\documentclass{aastex631}
\usepackage{rotating} %for the sideways table
\usepackage{xspace}
\newcommand{\numbursts}{828\xspace}
\newcommand{\numfbfits}{544\xspace}
\newcommand{\tns}{FRB~20220912A\xspace}
\newcommand{\dmvar}{$1.4 \pm 0.6$ pc cm$^{-3}$ yr$^{-1}$\xspace}
\begin{document}

% Title
\title{Radio Monitoring Campaign of Active Repeater FRB 20220912A with CHIME}
% Authors
\author[0000-0001-5002-0868]{Thomas C. Abbott}
\affiliation{Department of Physics, McGill University, 3600 rue University, Montr\'eal, QC H3A 2T8, Canada}
\affiliation{Trottier Space Institute, McGill University, 3550 rue University, Montr\'eal, QC H3A 2A7, Canada}

\author[0000-0002-8912-0732]{Aaron~B.~Pearlman}
\altaffiliation{NASA Hubble Fellow.}
\affiliation{MIT Kavli Institute for Astrophysics and Space Research, Massachusetts Institute of Technology, 77 Massachusetts Avenue, Cambridge, MA 02139, USA}
\affiliation{Department of Physics, Massachusetts Institute of Technology, 77 Massachusetts Avenue, Cambridge, MA 02139, USA}
\affiliation{Department of Physics, McGill University, 3600 rue University, Montr\'eal, QC H3A 2T8, Canada}
\affiliation{Trottier Space Institute, McGill University, 3550 rue University, Montr\'eal, QC H3A 2A7, Canada}

\author[0000-0001-9345-0307]{Victoria M. Kaspi}
\affiliation{Department of Physics, McGill University, 3600 rue University, Montr\'eal, QC H3A 2T8, Canada}
\affiliation{Trottier Space Institute, McGill University, 3550 rue University, Montr\'eal, QC H3A 2A7, Canada}
\affiliation{School of Physics and Astronomy, Tel Aviv University, Tel Aviv 69978, Israel}
\author[0000-0002-8897-1973]{Ayush Pandhi}
\affiliation{Department of Physics, McGill University, 3600 rue University, Montr\'eal, QC H3A 2T8, Canada}
\affiliation{Trottier Space Institute, McGill University, 3550 rue University, Montr\'eal, QC H3A 2A7, Canada}

\author[0000-0002-1800-8233]{Charanjot Brar}
\affiliation{National Research Council of Canada, Herzberg Astronomy and Astrophysics, 5071 West Saanich Road, Victoria, BC V9E 2E7, Canada}

\author[0009-0007-0757-9800]{Alyssa Cassity}
\affiliation{Department of Physics and Astronomy, University of British Columbia, 6224 Agricultural Road, Vancouver, BC V6T 1Z1 Canada}

\author[0000-0001-6422-8125]{Amanda M. Cook}
\affiliation{Department of Physics, McGill University, 3600 rue University, Montr\'eal, QC H3A 2T8, Canada}
\affiliation{Trottier Space Institute, McGill University, 3550 rue University, Montr\'eal, QC H3A 2A7, Canada}
\affiliation{Anton Pannekoek Institute for Astronomy, University of Amsterdam, Science Park 904, 1098 XH Amsterdam, The Netherlands}

\author[0000-0002-8376-1563]{Alice P. Curtin}
\affiliation{Department of Physics, McGill University, 3600 rue University, Montr\'eal, QC H3A 2T8, Canada}
\affiliation{Trottier Space Institute, McGill University, 3550 rue University, Montr\'eal, QC H3A 2A7, Canada}
\affiliation{Anton Pannekoek Institute for Astronomy, University of Amsterdam, Science Park 904, 1098 XH Amsterdam, The Netherlands}

\author[0000-0001-8384-5049]{Emmanuel Fonseca}
\affiliation{Department of Physics and Astronomy, West Virginia University, PO Box 6315, Morgantown, WV 26506, USA}
\affiliation{Center for Gravitational Waves and Cosmology, West Virginia University, Chestnut Ridge Research Building, Morgantown, WV 26505, USA}

\author[0000-0002-3382-9558]{B. M. Gaensler}
\affiliation{Department of Astronomy and Astrophysics, University of California, Santa Cruz, 1156 High Street, Santa Cruz, CA 95060, USA}
\affiliation{David A.\ Dunlap Department of Astronomy and Astrophysics, 50 St. George Street, University of Toronto, ON M5S 3H4, Canada}

\author[0000-0003-1884-348X]{Deborah Good}
\affiliation{Department of Physics and Astronomy, University of Montana, 32 Campus Drive, Missoula, MT 59812, USA}

\author[0000-0003-2317-1446]{Jason W.T. Hessels}
\affiliation{Department of Physics, McGill University, 3600 rue University, Montr\'eal, QC H3A 2T8, Canada}
\affiliation{Trottier Space Institute, McGill University, 3550 rue University, Montr\'eal, QC H3A 2A7, Canada}
\affiliation{Anton Pannekoek Institute for Astronomy, University of Amsterdam, Science Park 904, 1098 XH Amsterdam, The Netherlands}
\affiliation{ASTRON, Netherlands Institute for Radio Astronomy, Oude Hoogeveensedijk 4, 7991 PD Dwingeloo, The Netherlands}

\author[0009-0004-4176-0062]{Afrokk Khan}
\affiliation{Department of Physics, McGill University, 3600 rue University, Montr\'eal, QC H3A 2T8, Canada}
\affiliation{Trottier Space Institute, McGill University, 3550 rue University, Montr\'eal, QC H3A 2A7, Canada}

\author[0000-0002-4209-7408]{Calvin Leung}
\affiliation{Miller Institute for Basic Research, Stanley Hall, Room 206B, Berkeley, CA 94720, USA}
\affiliation{Department of Astronomy, University of California, Berkeley, CA 94720, United States}

\author[0000-0002-7164-9507]{Robert Main}
\affiliation{Department of Physics, McGill University, 3600 rue University, Montr\'eal, QC H3A 2T8, Canada}
\affiliation{Trottier Space Institute, McGill University, 3550 rue University, Montr\'eal, QC H3A 2A7, Canada}

\author[0000-0001-7348-6900]{Ryan Mckinven}
\affiliation{Department of Physics, McGill University, 3600 rue University, Montr\'eal, QC H3A 2T8, Canada}
\affiliation{Trottier Space Institute, McGill University, 3550 rue University, Montr\'eal, QC H3A 2A7, Canada}

\author[0000-0001-8845-1225]{Bradley W. Meyers}
\affiliation{Australian SKA Regional Centre (AusSRC), Curtin University, Kent Street, Bentley, WA 6102, Australia}
\affiliation{International Centre for Radio Astronomy Research (ICRAR), Curtin University, Bentley WA 6102, Australia}

\author[0000-0003-0510-0740]{Kenzie Nimmo}
\affiliation{Center for Interdisciplinary Exploration and Research in Astronomy, Northwestern University, 1800 Sherman Avenue, Evanston, IL 60201, USA}

\author[0000-0002-0940-6563]{Mason Ng}
\affiliation{Department of Physics, McGill University, 3600 rue University, Montr\'eal, QC H3A 2T8, Canada}
\affiliation{Trottier Space Institute, McGill University, 3550 rue University, Montr\'eal, QC H3A 2A7, Canada}

\author[0000-0002-4795-697X]{Ziggy Pleunis}
\affiliation{Anton Pannekoek Institute for Astronomy, University of Amsterdam, Science Park 904, 1098 XH Amsterdam, The Netherlands}
\affiliation{ASTRON, Netherlands Institute for Radio Astronomy, Oude Hoogeveensedijk 4, 7991 PD Dwingeloo, The Netherlands}

\author[0000-0002-7374-7119]{Paul Scholz}
\affiliation{Department of Physics and Astronomy, York University, 4700 Keele Street, Toronto, ON MJ3 1P3, Canada}

\author[0000-0002-4823-1946]{Vishwangi Shah}
\affiliation{Department of Physics, McGill University, 3600 rue University, Montr\'eal, QC H3A 2T8, Canada}
\affiliation{Trottier Space Institute, McGill University, 3550 rue University, Montr\'eal, QC H3A 2A7, Canada}

\author[0000-0002-6823-2073]{Kaitlyn Shin}
\affiliation{Division of Physics, Mathematics, and Astronomy, California Institute of Technology, Pasadena, CA 91125, USA}

% Abstract
\begin{abstract}
\tns is a highly active repeating fast radio burst (FRB) source, discovered by the Canadian Hydrogen Intensity Mapping Experiment (CHIME) using its real-time FRB detection system (CHIME/FRB). Here, we present results from a radio monitoring campaign of \tns using CHIME, including $\sim$200 hours of data collected by CHIME/Pulsar, spanning 1.5 years following the source’s discovery. We present an analysis of a sample of \numbursts CHIME-detected bursts from \tns, in the 400--800\,MHz radio frequency band. The source remains highly active for $\sim$10~weeks and has a bimodal wait-time distribution with peaks at 160$^{+120}_{-70}$\,ms and 306$^{+14}_{-13}$\,s. Assuming a radio efficiency factor of $10^{-4}$ and a beaming angle of $0.1$, we estimate the total emitted energy from the source over the entire observing campaign to be $2 \times 10^{43}$\,ergs. We report a 2.3$\sigma$ detection of a linear increase in the DM of \dmvar, with no significant trend in rotation measure (with a 3$\sigma$ upper limit of 13.4\,rad\,m$^{-2}$ yr$^{-1}$). We contrast our findings with other active repeaters, which exhibit different DM and RM evolution to indicate that \tns may reside in a unique local environment.
\end{abstract}

% Keywords

% Introduction
\section{Introduction}\label{sec: Introduction}
Fast radio bursts (FRBs) are energetic, short duration ($\mu$s -- ms) radio pulses primarily of extragalactic origin (for a review, see \citealt{frb_review_2019}). A small percentage ($\sim$ 4\%) of FRB sources have been observed to emit multiple bursts (Cook et al., In prep.). These repeating FRB sources show a wide variety of burst rates, with some sources detected only twice in hundreds of hours of exposure, while a handful have observed burst rates exceeding several hundred bursts per hour from a variety of telescopes (e.g., \citealt{gajjar_R1}, \citealt{FAST_R1_ATEL}, \citealt{lanman_2022_frb_20201124A}, \citealt{jahns_2022_frb_20121102a}, \citealt{nimmo_frb20200120E}, \citealt{shin_2026}). Sources exhibiting high burst rates, sometimes referred to as ``hyperactive", provide valuable clues to the nature of FRB progenitors and their environments. Long-term observations of highly active repeaters are valuable for tracking the evolution of emission properties and line-of-sight effects from local, interstellar, and intergalactic media, helping to constrain FRB progenitor models and assess the diversity of FRB source environments. These studies are complementary to other approaches which aim to understand FRB progenitors by localizing FRBs to specific hosts and galactic neighborhoods (e.g., 
\citealt{doi:10.1126/science.aaw5903}, \citealt{2024NatCo..15.7454Z},
\citealt{2025ApJS..280....6C},
\citealt{10.1093/mnras/staf2144}). 

\tns is a hyperactive repeating FRB that was discovered in 2022 \citep{r117_discovery_atel} by the Canadian Hydrogen Intensity Mapping Experiment Fast Radio Burst instrument (CHIME/FRB; \citealt{chimefrb_overview}). Shortly after the discovery, the Deep Synoptic Array (DSA) localized the source to a host galaxy, PSO J347.2702+48.7066, at a redshift of $z = 0.0771$ \citep{r117_host_galaxy}. Using the European VLBI Network (EVN), \cite{evn_localization} further refined the localization to within $\sim$5 mas (corresponding to a transverse scale of approximately 18 parsecs). For several weeks after the source's discovery, an assortment of radio observatories performed follow-up observations from 111 to 1700~MHz. The Five-hundred meter Aperture Spherical Telescope (FAST) detected burst rates of up to 390 bursts hr$^{-1}$ and showed that the estimated total emitted energy (assuming a radio efficiency factor of 10$^{-4}$) may challenge some FRB progenitor models involving magnetars (\citealt{fast_20201124A_extreme_activity}, \citealt{lpa_fast}, \citealt{zhang_2023_fast_observations_r117}). Through $\sim 1500$ hours of observation with a suite of four European radio telescopes, \cite{omar-2026-max-energy} found an exponential cut-off of the power-law slope of the spectral energy density at the high end of the source's burst energy distribution ($\sim 2 \times 10^{32}$ ergs Hz$^{-1}$ and $10^{32}$ ergs Hz$^{-1}$, respectively), providing insight on the maximum energetics of FRB sources. \cite{gmrt_long_term_campaign} find a consistent cut-off in the power-law slope using low-frequency uGMRT observations over a timespan of nearly two years. Finally, \cite{nancay} used $1.2 - 1.7$ GHz observations with the Nançay Radio Telescope (NRT) and revealed hints of an increasing dispersion measure (DM) when the source's activity rate declined, but could not confirm this due to covariance with burst morphology changes. The prolonged high activity rate from this source and the observations reported in previous works motivated our long-term follow-up campaign using the CHIME/Pulsar instrument, reported on here.

Located at the National Research Council of Canada’s Dominion Radio Astrophysical Observatory (DRAO) near Penticton, British Columbia, Canada, CHIME is a transit radio telescope with several processing backends specialized for a variety of scientific goals. The CHIME/FRB backend offers 1024 static digital beams spanning a $\sim100^\circ \times 2^\circ$ North-South strip of the sky at the meridian. Due to its large field-of-view ($\sim$ 200 deg$^2$), high sensitivity, and daily coverage of the Northern sky, CHIME/FRB is a powerful FRB detector (e.g., \citealt{chimefrb_cat1}, \citealt{chimefrb_cat2}). For bursts exceeding a SNR detection threshold of 12, the CHIME/FRB backend also records raw voltage data, making polarimetric studies possible \citep{chimefrb_basecat}. 

\begin{figure}[t]
    \centering
    \includegraphics[width=0.9\linewidth]{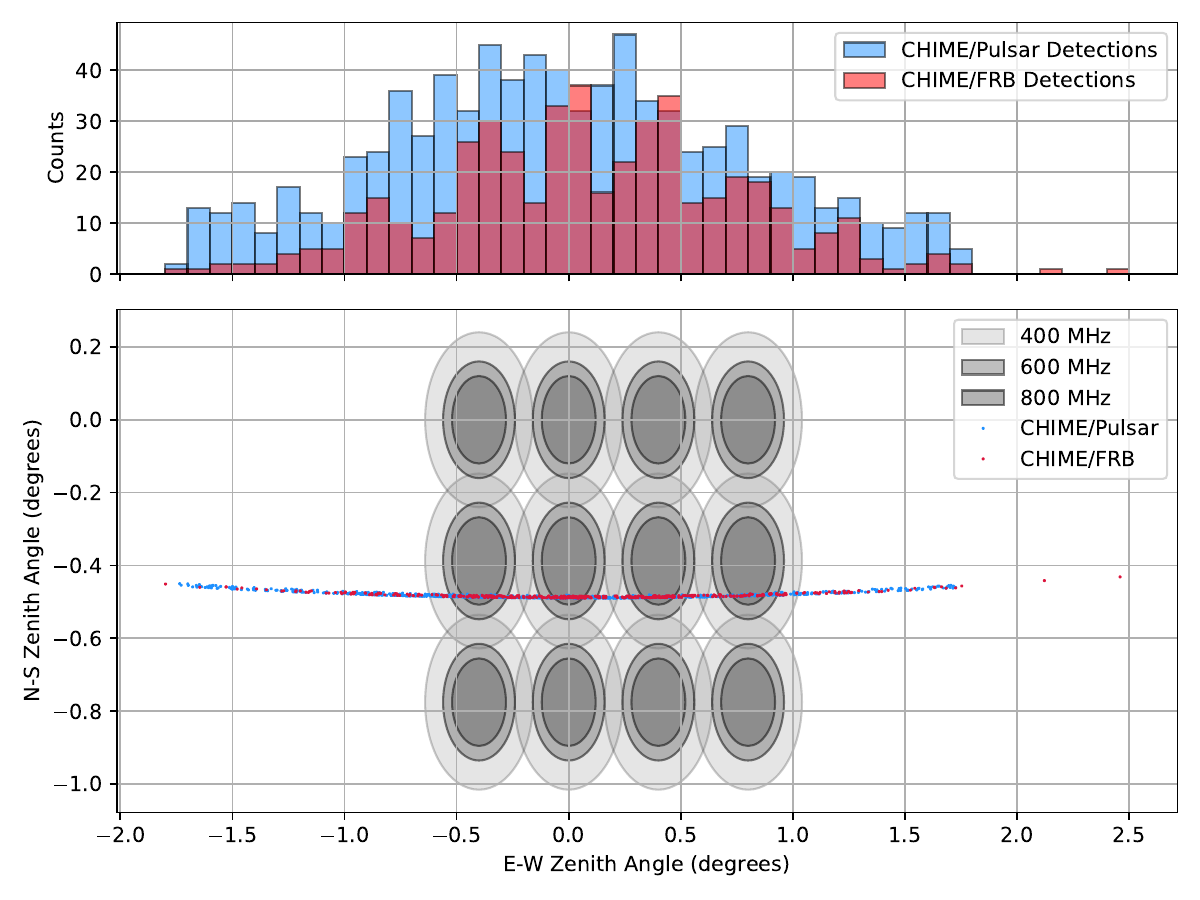}
    \caption{Comparison of CHIME/FRB detections and CHIME/Pulsar detections from this study over-plotted on a subset of 12 CHIME/FRB beams.  \textit{Upper Panel:} A histogram (binned every 0.1 degrees) showing the number of detections from each of the two CHIME backends. \textit{Lower Panel:} A subset of the CHIME/FRB beam grid projected onto the sky showing each beam size at $400$, $600$, and $800$ MHz in gray. Points where CHIME/FRB made a detection are shown as red dots and points where CHIME/Pulsar made a detection are shown as blue dots. Both panels show detections made between 2022 September 15 and 2024 June 18.}
    \label{fig: beam_plot}
\end{figure}

CHIME/Pulsar \citep{chimepsr_overview} is designed for high-time resolution, precision pulsar observations. For in-depth studies of single repeating sources, the CHIME/Pulsar instrument offers several advantages over CHIME/FRB. First, CHIME/Pulsar produces ten digitally formed tracking beams, which can each be centered on a source throughout its transit. This eliminates the frequency-dependent response of the static beams of CHIME/FRB, and is particularly beneficial for sources that transit through regions in the CHIME/FRB beam grid that have reduced sensitivity (see Figure \ref{fig: beam_plot}). Furthermore, CHIME/Pulsar is capable of recording high-time resolution, coherently de-dispersed filterbank data, enabling 40.96 $\mu$s single pulse searches for temporally narrow FRBs from repeating sources with known DM \citep{chimepsr_overview}.

The high burst rate of \tns, combined with the long-term monitoring campaign performed by CHIME/Pulsar, has enabled us to conduct a novel study of the temporal evolution of the source's burst parameters over year-long timescales. 
In this paper, we present the results from 1.5 years of daily monitoring of \tns using the CHIME/Pulsar instrument and compliment them with polarimetric data from CHIME/FRB. In Section \ref{sec: Methods}, we describe the analysis pipeline used to search for and measure parameters of the bursts. In Section \ref{sec: Results}, we present the distributions of FRB properties from our sample, and how the properties evolve with time. In Section \ref{sec: Discussion}, we discuss our results. We conclude in Section \ref{sec: Conclusion}.

% Methods
\section{Methods}\label{sec: Methods}

\subsection{FRB Search}\label{subsec: Methods/FRB Search}
This analysis used the CHIME/Pulsar instrument’s search-mode data which was recorded at a 40.96 $\mu$s time resolution and a 0.390625-MHz frequency resolution, covering $400-800$ MHz. For \tns, we recorded data on a daily basis for $21.2$ min across the transit centered on CHIME's meridian. The data were coherently de-dispersed to a DM of $219.456$ pc cm$^{-3}$ \citep{r117_discovery_atel} to correct for intra-channel dispersion and then saved in \texttt{filterbank} format. Prior to searching the data, the raw 8-bit \texttt{filterbank} data were converted to 32 bits and underwent a bandpass correction and baseline smoothing. Subsequently, 353 frequency channels with bad radio frequency interference (RFI) were masked and \texttt{PRESTO}'s \footnote{\url{https://www.ascl.net/1107.017}} \texttt{rfifind} method was applied to further clean the data. The masked data were downsampled in time by factors of $2 - 300$, corresponding to time resolution ranging between $81.92\ \mu$s and $24.6$ ms. The data were then incoherently de-dispersed (correcting for inter-channel dispersion) to a DM of $219.456$ pc cm$^{-3}$, and a matched-filter search was used to generate candidates. Events for which the de-dispersed time series reached or exceeded a \texttt{PRESTO} single-pulse search signal-to-noise ratio (SNR) of 5 were saved. The dynamic spectrum of each event where this occurred was plotted and visually inspected to verify which events are astrophysical in origin.

The search presented in this work occurred between 2022 September 15 and 2024 June 18, and led to 29,347 CHIME/Pulsar events that exceeded the SNR cutoff. Among those 29,347 events, \numbursts were promoted to FRB candidates through visual inspection, and underwent further processing. Table \ref{tab: burst parameters} lists the burst parameters for a portion of the detected bursts, ordered by arrival time. 

\subsection{Burst Parameter Pipeline}\label{subsec: Methods/Burst Parameter Pipeline}
The \numbursts human-verified candidates were extracted in 10-s intervals from the \texttt{filterbank} observation file, then masked using the already generated \texttt{rfifind} mask, as well as an additional inter-quartile range mitigation (IQRM) masking algorithm \citep{iqrm_mask}. The masked candidates were then input into \texttt{fitburst}, which determined the burst width, SNR-maximizing DM, and center frequency, using a least-squares fitting routine with a Gaussian profile \citep{fitburst}. Visually, we did not see any evidence for scattering from the dynamic spectra of the bursts, consistent with other studies at similar frequencies (\citealt{scattering_constraints}, \citealt{ovro}). Furthermore, when generating \texttt{fitburst} models for the ten narrowest bursts in the sample, models with no scattering were preferred. We thus report \texttt{fitburst} results for fits that include no scattering, and approximate the upper limit on the scattering timescale at 600 MHz to be 0.5 ms (the width of the narrowest bursts plus three times the uncertainty on the width). By reconstructing the \texttt{fitburst} model, the bandwidth of each burst was also determined. For each burst, at least two \texttt{fitburst} models were generated - a single-component fit, and a  multi-component fit. In some cases, the initial model parameters were manually adjusted to achieve a better fit. The best fit was selected by minimizing $\chi^2$, then fits and residual plots were validated visually. We obtained \texttt{fitburst} fits for \numfbfits of the bursts that had sufficient SNR and structure for the fitting method to converge. For each burst, we estimated the flux density, $S_\nu$, using the radiometer equation: 
\begin{equation}\label{eq: radiometer}
    S_\nu = \frac{\text{SNR} \cdot \text{SEFD}}{\sqrt{2 \cdot t_{obs} \cdot \text{BW}},}
\end{equation}
where $\text{SNR}$ is the signal-to-noise ratio, SEFD is the system equivalent flux density, $t_{obs}$ is the observation's time resolution (40.96 $\mu$s), and BW is the bandwidth of the total spectral extent of the burst across unmasked frequency channels. We calculated the SNR by summing the time series over the width of the pulse, subtracting the summed time series in an off-pulse region\footnote{We considered the off-pulse region to be the portion of the time series that was at least 5 full width half maxima (FWHMs) away from the burst center.} and then dividing by the off-pulse standard deviation. The SEFD was determined from calibration using standard cosmic radio sources \citep{chime_overview}. We correct for each burst's position within the CHIME primary beam pattern using the beam model described in \cite{chime_beammodel}. The burst fluences were determined by taking the larger value of the integrated flux over the full width half maximum (FWHM) and $2 \times $FWHM. Due to uncertainty in the SEFD, we conservatively used $50\%$ error bars on the resulting flux density values, as done by \cite{2025NatAs...9..111P}. Using our detection SNR threshold of 5 and assuming a characteristic burst width of 1 ms and SEFD of 55 K, we estimate the fluence threshold of our observations to be $0.92$ Jy ms.

We also report on polarimetry of \tns. While recording search-mode data, CHIME/Pulsar records total intensity only; however, for events with raw voltage data saved from the CHIME/FRB instrument, we obtain Faraday rotation measure (RM) values using the polarization processing pipeline presented by \cite{polarization_pipeline} and later updated by \cite{pandhi_2024}. In summary, we used the RM synthesis technique \citep{rm-synthesis}, which consists of testing trial Faraday depths and computing the corresponding average linearly polarized intensity in the Faraday dispersion function (FDF, assuming that the Faraday rotation seen in the bursts is Faraday simple)\footnote{We use RM synthesis instead of QU-fitting as QU-fitting faces some issues with cable delays that can lead to incorrect RM signs (\citealt{polarization_pipeline}, \citealt{pandhi_2024}). This is particularly impactful when the magnitude of the observed RM is close to zero.}.  For a polarized signal, the FDF is peaked, and the RM is obtained by taking the peak of the FDF value versus RM using a parabolic fit. The uncertainty in the RM is given by $\text{FWHM}/(2 \cdot L)$, where the FWHM of the FDF is determined by the parabolic fit and $L$ is the intensity of linear polarization.

% Results
\section{Results}\label{sec: Results}

\begin{table}
\centering
\begin{tabular}{lccccccc}
\hline\hline
TOA & SNR & DM & Width & Fluence & Bandwidth & $\alpha$ & d$\alpha$/d$\nu$  \\
(MJD) & & (pc cm$^{-3}$) & (ms) & (Jy ms) & (MHz) & & (Hz$^{-1}$) \\
\hline
59867.22872941(19) & 5.25 & 226.4(1) & 9.26(98) & 1.6(0.8) & 367(37) & 130(24) & $-$142(26) \\
59867.2287399112(95) & 5.60 & NA & 0.77(15) & 2.5(1.3) & 305(30) & $-$5.6(6) & 8.7(5) \\
59867.229201796(11) & 7.95 & 221.15(17) & 4.07(36) & 5.0(2.5) & 116(12) & 7.1(5) & $-$40(14) \\
59867.229217406(20) & 7.48 & 222.07(20) & 5.64(76) & 5.3(2.6) & 96.2(6) & 6.0(7) & $-$16.6(6) \\
59867.231601968(28) & 9.46 & 221.25(19) & 3.55(33) & 2.1(1.1) & 248(25) & 24.9(4) & $-$38.2(7) \\
59867.2329786985(95) & 5.15 & NA & 1.00(45) & 1.9(1.0) & 400(40) & 1.9(1) & $-$1.2(5) \\
59867.233022605(72) & 5.87 & 220.30(43) & 3.28(49) & 2.6(1.3) & 329(33) & 64(20) & $-$81(25) \\
59867.2333395978(95) & 5.30 & 219.571(41) & 4.51(56) & 2.1(1.0) & 353(35) & 94(17) & $-$109(20) \\
59868.2238143735(95) & 5.15 & NA & NA & 4.6(2.3) & NA & NA & NA \\
59868.225646473(32) & 6.01 & 219.99(19) & 1.17(19) & 2.9(1.4) & 354(35) & 98(36) & $-$114(42) \\
59868.2269024807(95) & 5.21 & NA & NA & 1.4(0.7) & NA & NA & NA \\
59868.227766729(52) & 5.90 & 219.10(26) & 1.18(38) & 2.3(1.1) & 386(39) & 249(75) & $-$214(64) \\
59868.2301085754(95) & 5.54 & 219.577(38) & 2.43(41) & 3.5(1.8) & 249(25) & 56(15) & $-$135(33) \\
59868.2314913584(31) & 5.86 & 219.631(33) & 0.87(16) & 3.1(1.6) & 400(40) & 0.2(9) & $-$6.2(8) \\
59868.2353383899(95) & 5.15 & NA & NA & 2.3(1.1) & NA & NA & NA \\
59869.2201067067(95) & 5.19 & NA & NA & 1.5(0.7) & NA & NA & NA \\
59869.223380503(95) & 5.09 & NA & NA & 1.8(0.9) & NA & NA & NA \\
59869.223541362(95) & 5.13 & NA & NA & 2.1(1.1) & NA & NA & NA \\
59869.2245944588(95) & 5.29 & 219.565(39) & 1.95(44) & 2.8(1.4) & 165(17) & 13.6(4) & $-$34(13) \\
59869.2246507067(95) & 5.14 & 219.4669(87) & 0.76(16) & 1.7(0.9) & 377(38) & 146(53) & $-$126(46) \\
59869.2271299586(95) & 5.48 & 219.532(28) & 3.38(42) & 1.4(0.7) & 362(36) & 88(32) & $-$77(28) \\
59869.227130619(23) & 6.55 & 219.63(19) & 2.36(49) & 2.8(1.4) & 202(20) & 19.0(8) & $-$38(13) \\
59869.228360321(37) & 6.84 & 220.10(19) & 1.46(19) & 3.6(1.8) & 359(36) & 87(31) & $-$82(30) \\
59869.230276790(27) & 6.35 & 218.09(24) & 7.0(5) & 2.3(1.1) & 363(36) & $-$13.1(4) & 18.0(5) \\
59870.2204642638(50) & 10.39 & 219.452(38) & 0.80(16) & 4.3(2.1) & 138(14) & 9.1(0) & $-$18.9(8) \\
59870.2218183218(95) & 5.56 & 219.470(11) & 0.73(21) & 2.2(1.1) & 303(30) & 35(14) & $-$39(15) \\
59870.2234252508(24) & 5.95 & 219.455(85) & 0.91(10) & 0.6(0.3) & 187(19) & $-$2.0(0) & $-$94(87) \\
59870.224423184(18) & 6.46 & 219.39(13) & 1.68(35) & 1.4(0.7) & 198(20) & 14.4(6) & $-$23.0(0) \\
59870.225071077(11) & 5.76 & 219.746(81) & 1.59(21) & 0.7(0.3) & 172(17) & 11.8(4) & $-$20.8(7) \\
59870.2266841309(95) & 5.18 & 219.273(74) & 3.13(43) & 2.0(1.0) & 154(15) & 45(20) & $-$180(76) \\
59871.218614453(40) & 9.29 & 221.15(24) & 2.79(21) & 6.0(3.0) & 356(36) & 118(18) & $-$150(23) \\
59871.2216349269(95) & 5.61 & 219.737(79) & 2.53(31) & 1.3(0.7) & 126(13) & 38(11) & $-$246(65) \\
59871.2219367756(20) & 15.91 & 219.584(22) & 0.736(39) & 1.7(0.8) & 203(20) & 29.7(3) & $-$91.0(7) \\
59872.2114596876(95) & 5.03 & NA & NA & 2.4(1.2) & NA & NA & NA \\
59872.2120473385(38) & 7.69 & 219.597(54) & 1.35(19) & 3.9(1.9) & 131(13) & 9.2(1) & $-$50(16) \\
59872.2121813687(69) & 9.22 & 219.33(14) & 3.08(32) & 12.7(6.3) & 190(19) & 14.8(7) & $-$106(26) \\
59872.214154150(78) & 8.81 & 219.76(39) & 2.34(17) & 10.1(5.0) & 389(39) & 367(68) & $-$348(65) \\
59872.216858754(19) & 8.68 & 219.90(12) & 2.92(33) & 3.8(1.9) & 232(23) & 18.1(1) & $-$24.7(5) \\
59872.2170304145(76) & 5.83 & 219.29(11) & 0.60(14) & 1.4(0.7) & 166(17) & 41(19) & $-$157(65) \\
59872.2190356872(27) & 6.15 & 219.379(34) & 0.630(70) & 9.1(4.5) & 172(17) & 19.0(7) & $-$75(17) \\
59872.219036218(17) & 9.04 & 220.69(11) & 3.36(27) & 0.8(0.4) & 217(22) & 15.5(7) & $-$21.1(4) \\
\hline
\end{tabular}
\caption{Table of \tns burst properties. We present a subset of the first 60 detected \tns bursts sorted in order of detection date. TOA is the topocentric time of arrival at CHIME in MJD format referenced to 800 MHz, SNR is the \texttt{PRESTO} detection \citep{presto} single pulse search signal-to-noise ratio, DM is the dispersion measure of the burst in pc cm$^{-3}$ (for bursts where there is no \texttt{fitburst} DM fit available, no DM is reported, however, note that the search DM used to initially detect all bursts was $219.456$ pc cm$^{-3}$). We use a dispersion constant of $1/(2.41 \times 10^{-4})$ GHz$^2$ pc$^{-1}$ cm$^3$ s. Also shown are the burst widths, bandwidths, the spectral indices ($\alpha$), and spectral running (d$\alpha$/d$\nu$) parameters measured by \texttt{fitburst}, if available. For some events, the DM or arrival time has no reported uncertainty despite a \texttt{fitburst} model being available. This is because those parameters were held fixed to perform the fit and are therefore not measured parameters.}\label{tab: burst parameters}
\end{table}

\subsection{Burst Activity Over Time}\label{subsec: Results/Burst Activity Over Time}
\begin{figure}[ht]
    \centering
    \includegraphics[width=\linewidth]{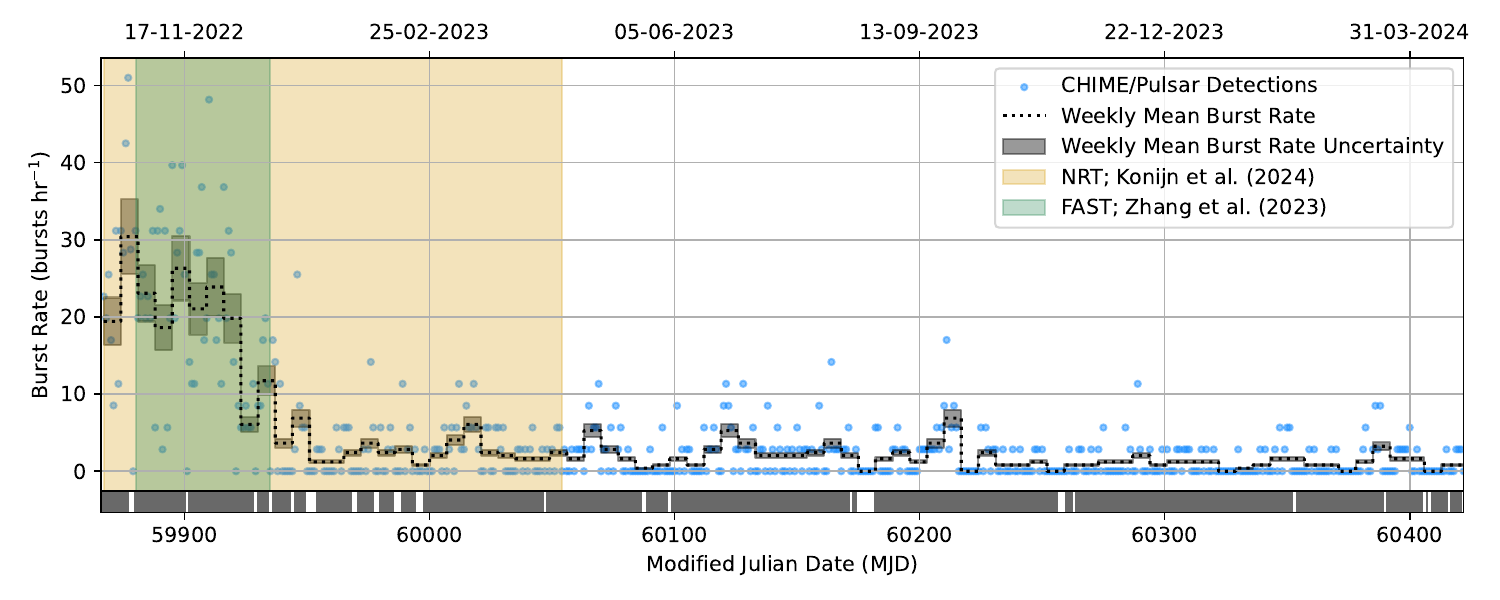}
    \caption{Burst rate versus time for \tns. The blue points show the mean burst rate from each CHIME/Pulsar observation. The black dotted line shows the mean burst rate observed by CHIME/Pulsar over week-long intervals with uncertainties. The highlighted regions show the periods where FAST (green) and NRT (yellow) observed the source. The gray bar along the bottom of the plot shows days on which CHIME/Pulsar tracked the source for a full 21.2-minute transit (gray) versus days where CHIME/Pulsar was down or did not track the source for the nominal 21.2-minute transit (white).}
    \label{fig: burst rate}
\end{figure}

The source's activity over time as observed from CHIME/Pulsar is shown in Figure \ref{fig: burst rate}. We determine the burst rate by dividing the number of bursts detected in a transit by the transit time in hours. In total, \numbursts bursts were detected from the source in 201.2 hours of observation, implying a mean burst rate of (4.12 $\pm$ 0.07) bursts hr$^{-1}$ at a fluence threshold of $0.92$ Jy ms. The peak observed burst rate over a 21 minute transit was 52 $\pm$ 2 bursts hr$^{-1}$ and it occurred on 2022 October 25 (corresponding to MJD 59877). In Figure \ref{fig: burst rate}, we also plot the mean burst rate with uncertainty over week-long intervals to highlight the variation in burst rate over time. We find that the mean weekly burst rate changes significantly with time; in particular, the source undergoes high activity in the first 10 weeks of observation, after which; most observations result in non-detections. On some individual days, the burst rate rises well above the mean; however after the initial 10 weeks of high activity, the weekly mean burst rate is below or consistent with the overall mean burst rate. Furthermore, the source position is observed by CHIME/FRB with daily cadence and had accumulated $\sim 226.7$ hours of exposure prior to being detected. This implies a $3\sigma$ upper limit on the burst rate prior to detection of $0.026$ bursts hr$^{-1}$. 

In comparison to FAST and NRT during their respective observing windows (see Figure \ref{fig: burst rate} for the observing times), NRT observed a peak mean burst rate of $75^{+10}_{-9}$ bursts hr$^{-1}$ while FAST observed $\sim$124 bursts hr$^{-1}$. Differences in burst rates can be attributed to the high source activity at the relevant observing epoch and the greater telescope sensitivities of FAST and NRT, but may also be influenced by their higher observing frequencies.

\subsection{Waiting Time Distribution}\label{subsec: Results/Waiting Time Distribution}
\begin{figure}
    \centering
    \includegraphics[width=\linewidth]{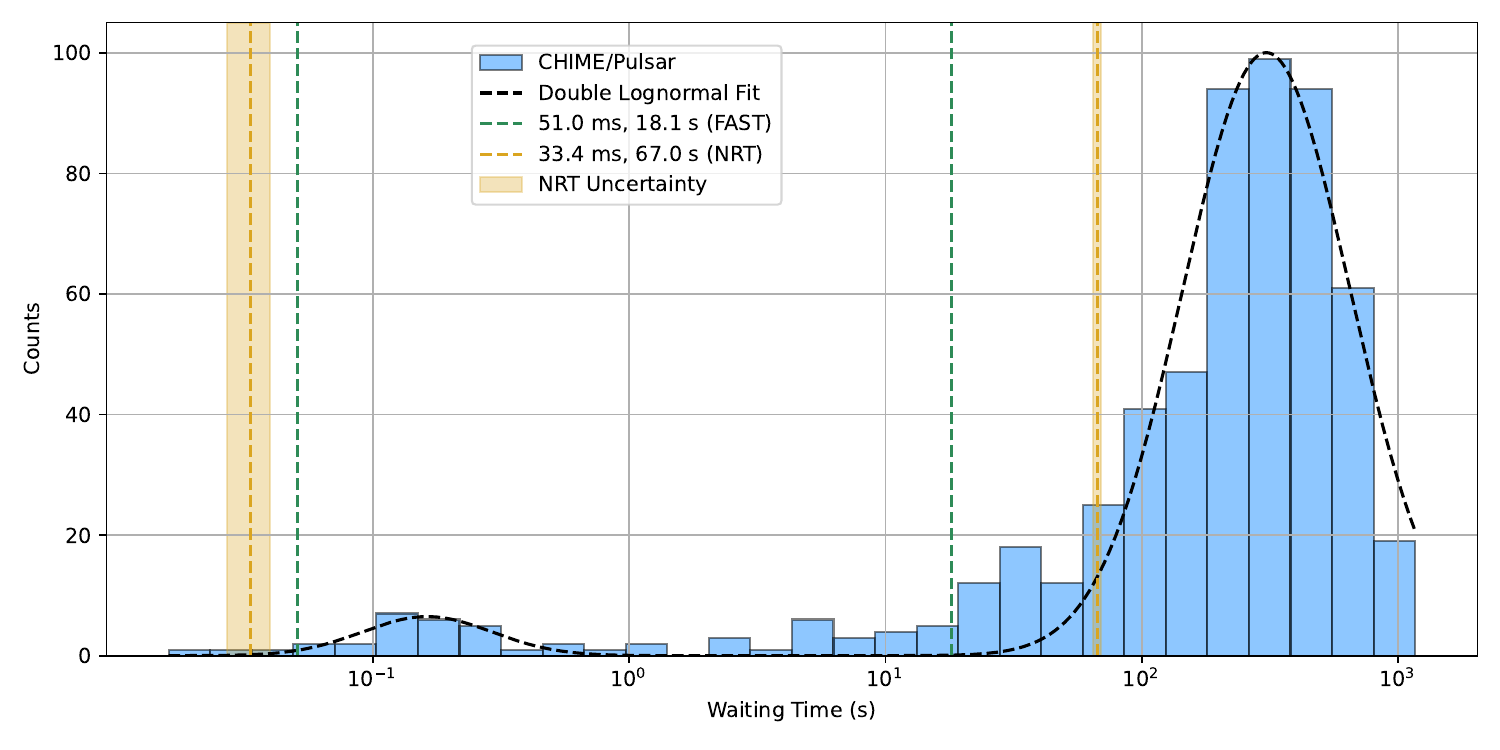}
    \caption{Waiting time distribution of \tns. The blue distribution shows the wait-times across the entire CHIME/Pulsar observing campaign (400 MHz $-$ 800 MHz). The yellow dashed lines and highlighted region show the mean waiting times observed by the NRT (1.2 $-$ 1.7 GHz) with uncertainties highlighted in yellow \citep{nancay} and the green dashed lines and shaded region show those of FAST (1 $-$ 1.5 GHz) with uncertainties not reported \citep{zhang_2023_fast_observations_r117}. The black dashed line shows the double lognormal best-fit used to measure the waiting time modes of the CHIME/Pulsar distribution and corresponding uncertainties.}
    \label{fig: wait-time distribution}
\end{figure}

The waiting time between bursts, defined in this study as the difference in time between consecutive bursts or consecutive sub-components between which the intensity time series reaches the noise floor, may provide clues towards the nature of FRB sources. In the case of \tns, \cite{zhang_2023_fast_observations_r117} and \cite{nancay} determine the waiting time distribution as observed by FAST and NRT, respectively, and find that it is bimodal with the shorter mode centered around tens of milliseconds (peaking at 51.0 ms and 33.4 ms, respectively)
and the longer mode centered around tens of seconds (peaking at 16.0 s and 67.03 s, respectively). In Figure \ref{fig: wait-time distribution}, we present the waiting time distribution of \tns as observed by CHIME/Pulsar over the entire observing campaign. We have made no correction for observational biases such as sensitivity and observing frequency in making this plot, consistent with other studies by different telescopes.  Using a double lognormal fit, we find that the shorter mode observed with CHIME/Pulsar occurs at 160$^{+120}_{-70}$ ms and the longer more prominent mode occurs at 306$^{+14}_{-13}$ s. The CHIME/Pulsar derived waiting time modes are not consistent with those found in the FAST and NRT studies, however that is not unexpected as these telescopes observed the source during a more active epoch and have different fluence thresholds and observing frequencies. We discuss this further in Section \ref{subsec: Discussion/Activity Rate and Waiting Time Distribution}.

\subsection{Isotropic Equivalent Energies}\label{subsec: Results/Isotropic Equivalent Energies}

\begin{figure}
    \centering
    \includegraphics[width=\linewidth]{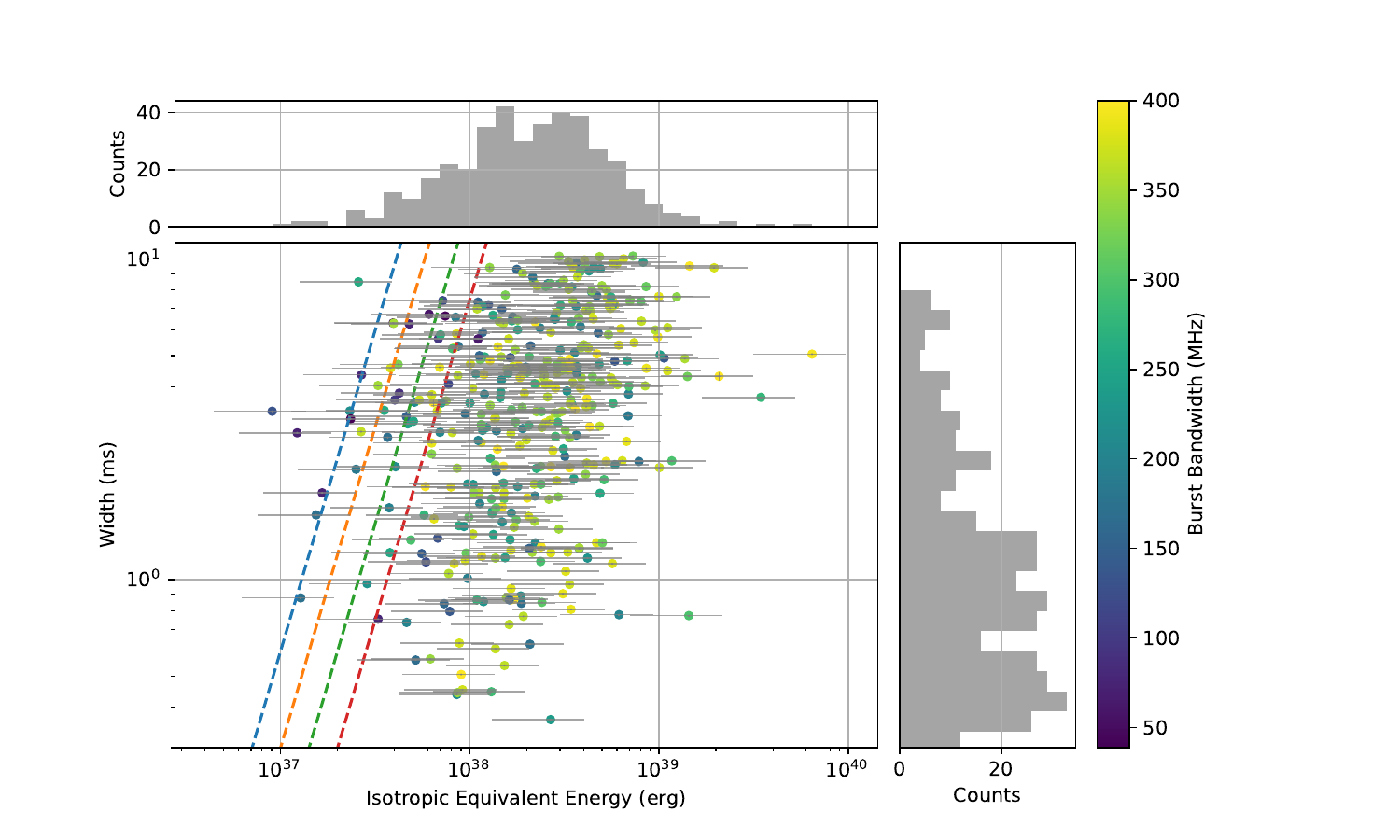}
    \caption{Distribution of burst widths versus isotropic equivalent energy for bursts with $F>0.92$ Jy ms. The scatter plot shows burst widths in milliseconds versus isotropic equivalent energy plotted logarithmically, where points are colored by each burst's measured bandwidth within the CHIME band. We also show four colored dashed lines representing a SNR threshold of 5 at various burst bandwidths ($50$ MHz in blue, $100$ MHz in orange, $200$ MHz in green, and $400$ MHz in red). Note that no effort has been made to correct for instrumental bias in these plots, therefore correlations between bandwidth and isotropic equivalent energy may be instrumental.  The top panel and right panel show the distributions of isotropic equivalent energies and widths, respectively.}
    \label{fig: iee-width-distribution}
\end{figure}

In Figure \ref{fig: iee-width-distribution}, we show the isotropic equivalent energy of each burst versus its width, for bursts for which the fluence is well measured (i.e. $F> 0.92$ Jy ms).  Note that no effort has been made here to correct for instrumental selection bias, which manifests at low-energy (fluence) where our sensitivity drops off, particularly for wider bursts.  To determine the isotropic equivalent energy of each burst, we use Equation \ref{eq: isotropic_equivalent_energy}:

\begin{equation}\label{eq: isotropic_equivalent_energy}
    E = \frac{4\pi \cdot F \cdot \text{BW} \cdot D_L^2}{(1 + z)^2},
\end{equation}
where $E$ is the isotropic equivalent energy (in ergs), $F$ is the fluence, BW is the bandwidth, $D_L$ is the luminosity distance, and $z$ is the redshift to the source ($D_L = 362.4$ Mpc and $z = 0.0771$ for \tns; \citealt{r117_host_galaxy}). 

We do not see distinct clustering of bursts in Figure \ref{fig: iee-width-distribution} between the higher and lower energy bursts, nor any distinction in their widths. The slight positive correlation between width and isotropic equivalent energy is naturally expected for bursts with the same SNR threshold. We note that this is in contrast to the reported findings of \cite{arecibo_heb_and_leb} who found that their sample of bursts from FRB 20121102A detected by Arecibo grouped into a high energy burst cluster ($10^{39} - 10^{40}$ ergs), and a low-energy burst cluster ($10^{37} - 10^{38}$ ergs), where the low-energy bursts tended to have longer duration and narrower bandwidths, possibly related to the observed differences between one-off and repeating bursts, where repeat bursts tend to be temporally wider with narrower bandwidths \citep{RN3}.

\subsection{Evolution of the Dispersion Measure and Rotation Measure}\label{subsec: Results/Evolution of the Dispersion Measure and Rotation Measure}
We measure DM using \texttt{fitburst}, which de-disperses the data to the DM that maximizes the SNR ratio. The DMs measured using this method can be biased by other sources of drift in the burst's dynamic spectra that may appear like dispersion (\citealt{2019ApJ...876L..23H}, \citealt{r117_dm_paper}). As a result, we visually verify each DM and focus our DM analysis on the DM measurements from bright, temporally narrow bursts, on which effects of drift are less significant. Based on the findings of \cite{r117_dm_paper}, we present the DM measurements of the narrow (width $\leq$ 2 ms) bright (SNR $\geq$ 8) bursts in the sample. Some examples of these bursts and their \texttt{fitburst} fits are shown in Appendix \ref{sec: appendix/fitburst_fits}. In Figure \ref{fig: dm varation},  By performing a linear regression fit, we find that the DM of these bursts increases linearly with time, with a slope of \dmvar. The slope is inconsistent with zero at roughly the 98\% confidence level ($\approxeq 2.3\sigma$). Furthermore, we note that all 43 bursts in Figure \ref{fig: dm varation} detected after MJD 59931 (corresponding to 2022 December 18) deviate by more than $3\sigma$ from the detection DM of 219.456 pc cm$^{-3}$. We caution that the DM fit may not have strong predictive power over long time scales.

Since the CHIME/FRB and CHIME/Pulsar instruments operate in parallel, there are 252 \tns events detected by both systems. 
Among those events, 222 meet the criteria for the CHIME/FRB system to store raw voltage data. The raw voltage, or baseband data, contain polarization information enabling RMs to be obtained as described in Section \ref{subsec: Methods/Burst Parameter Pipeline}. RM values for \tns are presented in the lower panel of Figure \ref{fig: dm varation}. The observed RM varies about zero across the entire observing campaign. The slope of the line of best-fit to the RM is $3.8 \pm 3.2$ rad m$^{-2}$ yr$^{-1}$, indicating a 3$\sigma$ upper limit on $\frac{d}{dt}(\text{RM})$ of 13.4 rad m$^{-2}$ yr$^{-1}$. Similarly to \cite{zhang_2023_fast_observations_r117}, nearly all bursts in our sample with raw voltage data are found to be nearly 100\% linearly polarized, with no clear evidence for depolarization.

\begin{figure}
    \centering
    \includegraphics[width=\linewidth]{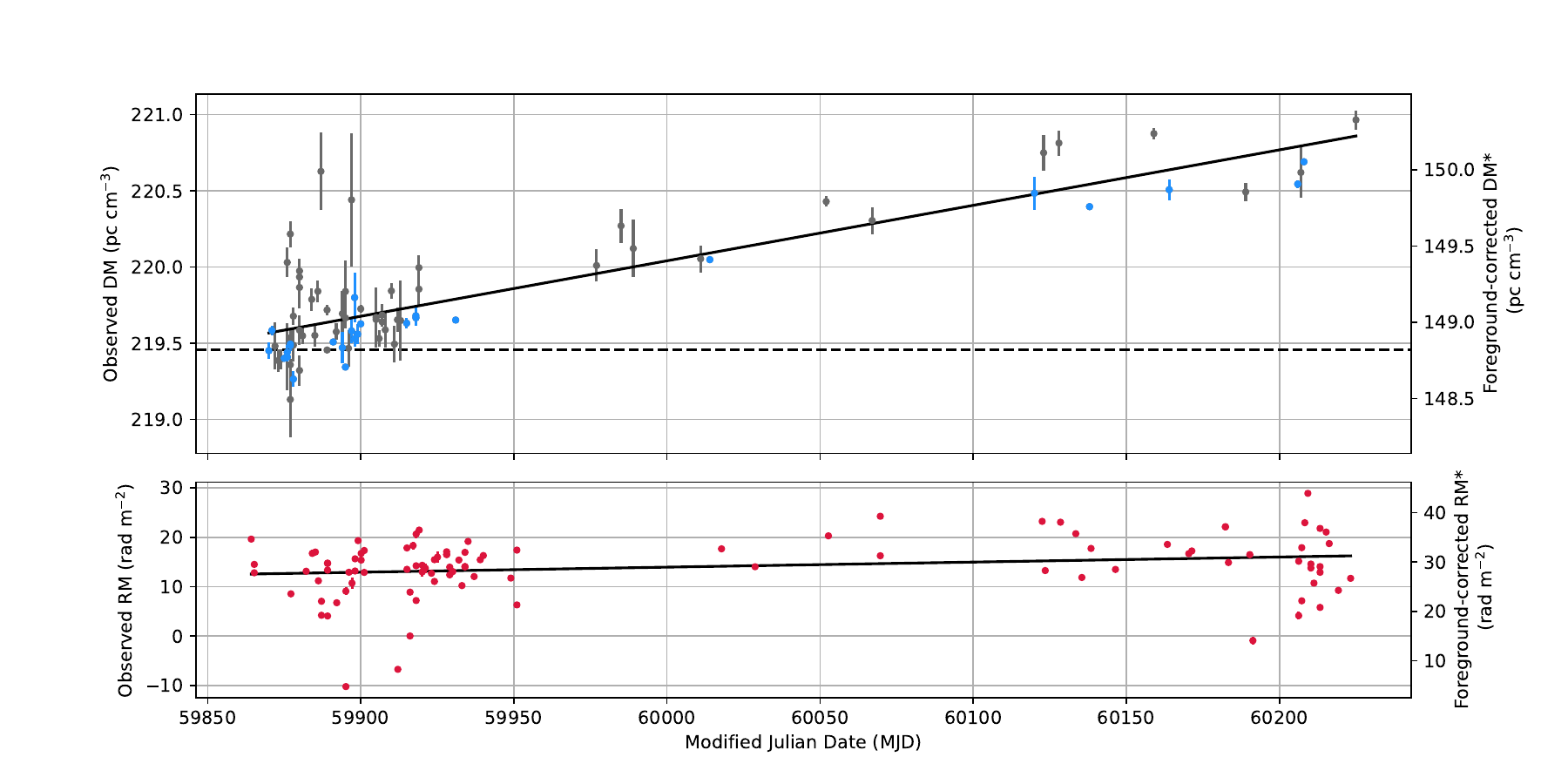}
    \caption{Temporal evolution of the DM and RM. \textit{Upper Panel:} DM in units of pc cm$^{-3}$ versus MJD. Blue dots show measurements for which a burst's signal to noise ratio exceeded 10 and temporal width was less than 1 ms. Gray dots show measurements for which burst's signal to noise ratio exceeded 8 and temporal width was less than 2 ms. The DM uncertainties shown are determined by \texttt{fitburst} and represent the measurement uncertainties. The solid black line shows the best-fit line to these values, which has a slope of \dmvar. The dashed black line shows the detection DM of 219.456 pc cm$^{-3}$ \citep{r117_discovery_atel}. The secondary vertical axis shows the DM after subtracting the estimated Galactic DM contribution \citep{ne2001}. \textit{Lower Panel:} RMs of co-detected events with baseband data recorded by CHIME/FRB (red) in units of rad m$^{-2}$ versus MJD. The RM errors are determined by the pipeline described by \cite{pandhi_2024} and represent measurement uncertainties. The secondary vertical axis shows the RM after subtracting the estimated Galactic RM contribution \citep{hutschenreuter2022}. *Uncertainty on the foreground-corrected values for the secondary axes are not shown.}
    \label{fig: dm varation}
\end{figure}

% Discussion
\section{Discussion}\label{sec: Discussion}

\subsection{Activity Rate and Waiting Time Distribution
}\label{subsec: Discussion/Activity Rate and Waiting Time Distribution}
In Section \ref{subsec: Results/Burst Activity Over Time}, we showed that the source's activity rate changes significantly with time. In particular, Figure \ref{fig: burst rate} showed that activity peaked in the 10 weeks following the source's discovery, after which most observations were non-detections. Significant time variation in repetition rate has been observed in other repeaters.  For example, \cite{lanman_2022_frb_20201124A} presented a variation in the repetition rate of FRB 20201124A. The source was undetected in 41.42 hours of exposure using CHIME/FRB before entering a state of high activity (between 92 and 201 bursts per day) in the month of April 2021, after which it became mostly inactive. 
FRB 20240114A is another source that appeared suddenly in early 2024 with extremely high burst activity at many radio frequencies \citep[see][and references therein]{shin_2026} that has mostly waned since.

One model for sources that exhibit sudden surges in activity for month-long timescales followed by a burst rate decline is the recent formation of young magnetars (see \citealt{beaming_factor}, for a review). These models can be tested in a number of ways. For example, most young magnetar models predict contemporaneous high energy bursts. For \tns, \cite{amanda_xray_followup} performed contemporaneous observations at X-ray and radio energies while the source was highly active, but detected no high energy emission, and set an upper limit on the X-ray to radio fluence ratio of only a factor of $\sim$3 higher than that observed from the 2020 FRB-like burst from the Galactic magnetar SGR 1935+2154 \citep{chimefrb_sgr}. Additionally, young magnetar models generally predict a monotonic decline in the burst rate, thus, a strong reactivation of a hyperactive repeating source would disfavor these models. Continued monitoring of repeating FRBs in hyperactive states will be helpful for constraining models. 

The bimodal waiting time distribution that we find for \tns presented in Figure \ref{fig: wait-time distribution} is qualitatively similar to those found by \cite{zhang_2023_fast_observations_r117} and \cite{nancay} as well as studies of other hyperactive repeating sources \citep[e.g.][]{xnc+22,nimmo_frb20200120E}. However, the two peaks that we detect are at larger waiting times than their counterparts seen in the other studies of \tns. This is influenced by the fact that the NRT and FAST have lower fluence thresholds, which leads to more detected bursts, and therefore shorter waiting times. However, it may be due to a combination of the different burst rates during our respective observations, observing frequencies, and observation durations. Our criteria to measure a wait-time is to measure the difference in TOA between bursts or burst sub-components separated by a point where the intensity reaches the noise level, as done in \cite{nancay}, however, we caution that we are unsure if this is the same criteria used in \cite{zhang_2023_fast_observations_r117} and this may affect our comparison.

For the shorter waiting time mode, which \cite{nancay} suggest might represent a physical scale size for the emission region, is mostly a result of time differences between burst sub-components. Telescopes with finer time resolution and greater sensitivity will observe shorter waiting time mode, as they are more capable of detecting faint, temporally narrow bursts. The time resolutions of the CHIME, NRT, and FAST studies are 41.92 $\mu$s, 16 $\mu$s, and 49.152 $\mu$s, respectively. This suggests that the NRT search is better suited than the CHIME/Pulsar or FAST searches to identify temporally narrow bursts. The difference seen between CHIME and FAST's shorter waiting time peak is likely a result of differences in sensitivity, where FAST is probing fainter bursts and burst sub-components, therefore shortening its observed short waiting time mode. We emphasize that these comparisons do not account for spectral variation in the number of components of FRBs from this source and the time at which the source was observed. The fact that comparing the shorter waiting time mode between studies depends on telescope sensitivity argues against physical interpretation of any one measured timescale.  Nevertheless the common preference for {\it some} shorter timescale is interesting and hints at underlying physics.

The larger waiting time mode has been shown, for several hyperactive repeaters, to be similar to the expected distribution from starquakes \citep{totani_magnetar_flares}. For this waiting time mode,  longer (both in individual observation time and total observation time) and more sensitive observations are thought to produce shorter modes. This is a result of more bursts being detected, and there being less influence from the finite observation time. While some of the falloff at long waiting time in our distribution could be due to our 21 minute transit time, correcting for that bias would only move the longer peak to longer timescales, farther from those of the other studies, and if anything would reinforce the bimodality. The collecting area of CHIME and the NRT is 8000 m$^2$ while FAST's collecting area is 196 000 m$^2$. The system temperature of CHIME, the NRT, and FAST is approximately 55 K, 20 K, and 35 K, respectively (\citealt{FAST_overview}, \citealt{chime_overview}, \citealt{nancay}). This suggests that the NRT is about 3 times as sensitive and FAST is about 39 times as sensitive as CHIME, matching the ordering seen in Figure \ref{fig: wait-time distribution}. The other factors that influence this difference which are not accounted for are the waiting time distribution's dependence on frequency, the total observing times, and the upper limit on individual observation durations.

\subsection{DM and RM Evolution}\label{subsec: Discussion/DM and RM Evolution}
The combination of long-term monitoring and high activity from \tns has enabled a unique study of its DM and RM evolution, especially using CHIME, as its low observing frequencies and large fractional bandwidth give a powerful lever arm on these chromatic effects. The linearly increasing DM of \dmvar suggests that there is a steadily increasing column density of ionized matter intervening the line of sight. 
There are different possibilities for the origin of this increase. Although it is unlikely to arise in the IGM where such a rapid change is not expected \citep[e.g.][]{frb_dm_variaton_causes}, it may arise in either the Milky Way, or the host galaxy -- whether in the interstellar medium (ISM) or in a region local to the source.

In the Milky Way, long-term monitoring campaigns of radio pulsars have provided information on time variations of Galactic DM. In an early study,
\cite{backer93} found DM variations that roughly correlated with DM and/or distance, but with maximal gradients of $<0.01$ pc cm$^{-3}$ yr$^{-1}$. In addition, in a 6-year study of 168 Galactic radio pulsars, \cite{petroff_psr_dm_variation} found a maximal DM variation of 0.14 pc cm$^{-3}$ yr$^{-1}$. Furthermore, the Galactic latitude of \tns is approximately $-10.8^\circ$, placing it away from the Galactic plane where most of the DM variation seen by \cite{petroff_psr_dm_variation} occurred. This strongly suggests that the DM variation we have detected for \tns is unlikely to be due to the Milky Way ISM. 

The source of the DM variation we observe could be in the FRB's host galaxy, if its ISM significantly differs from that of the Milky Way and/or if perhaps the source has a large proper motion. However, the relatively low inferred DM contribution (a 90\% confidence upper limit on the host DM of $53$ pc cm$^{-3}$) from the source's host galaxy PSO J347.2702+48.7066 \citep{r117_host_galaxy} suggests the host ISM contribution to the observed DM variation in \tns is analogously small. Similar conclusions have been made in other studies of DM variation in FRBs (e.g., \citealt{Hessels_2019}, \citealt{dm_variation_frb20180301a}, \citealt{dm_variation_frb20190520b}, \citealt{ayush_2026}). \cite{frb_dm_variaiton_causes} and \cite{piro_and_gaensler_2018} estimate expected DM magnitudes and signs of several sources of DM variations. There are several scenarios in which the DM variation is positive\footnote{See Section 2 of \cite{frb_dm_variaiton_causes} for the assumptions used to make the DM variation predictions shown.}, such as a local supernova remnant in its snow-plow phase ($\sim 0.01$ pc cm$^{-3}$ yr$^{-1}$), a growing H II region ($\sim 0.8$ pc cm$^{-3}$ yr$^{-1}$), or plasma lensing ($\sim 0.8$ pc cm$^{-3}$ yr$^{-1}$). We find no evidence for plasma lensing in the CHIME/Pulsar sample of bursts for this source, so given the magnitude of the DM variation in \tns,  it seems best  attributed to a growing H II region. The marginally larger measured DM variation (\dmvar) compared to the predicted DM variation for such a scenario could arise if the gas number density in the H II region is slightly higher or if the star within the H II region boosts the emission rate of the ionizing photons relative to the nominal values assumed in the model. An H II region might be observed as a persistent radio source (PRS), though \cite{dante_microshots} initially ruled out a PRS associated with \tns at milliarcsecond scales. Interestingly, \cite{ncrt20220912A} later found evidence for a PRS with a flux density of $240 \pm 36$ $\mu$Jy at $1.26$ GHz. Indeed, using the estimated 0.9 kpc offset of the PRS with the host galaxy's center, they suggest that this region is unlikely to originate from an active galactic nucleus, supporting an H II region hypothesis.

It is noteworthy that despite the DM varying, the observed RM remains close to zero and shows no evidence of long-term temporal evolution (with a $3 \sigma$ upper limit of $13.4$ rad m$^{-2}$ yr$^{-1}$). While both DM and RM depend on the integrated electron column density ($n_e$), RM additionally depends on the parallel component of the line of sight magnetic field ($B_{||}$). Consequently, variations in DM cause corresponding changes in RM if the varying plasma is magnetized (though these changes may not be proportional in amplitude). The estimated Galactic RM contribution along the line of sight is ($-$15 $\pm$ 11) rad m$^{-2}$ \citep{hutschenreuter2022}, which is within $2\sigma$ of our measured values but not strictly zero. Given uncertainties, and accounting for rest-frame scaling, we cannot rule out a small host-galaxy RM contribution. As argued above, the change in DM is likely local to the source, so the small RM strongly suggests the FRB environment is not strongly magnetized \citep[see also][]{weakly_magnetized_env}. Considering the scenario where the DM change occurs in the local environment of the source, we place a $3\sigma$ upper limit on the magnetic field strength of the environment of $8.8$ $\mu$G (applying Equation 2 from \citealt{b-field-equation}). 

In a sample of 40 repeating FRB sources with four or more detected bursts, Cook et al. (In prep.) use ms-resolution CHIME/FRB data to show that \tns is among only a handful of sources to display DM variation. By contrast, the repeating FRB 20121102A was shown to exhibit an RM in excess of $10^5$ rad m$^{-2}$ \citep{michilli18}, with short-term variations of order $10^3$ rad m$^{-2}$ \citep{hilmarsson21}, and a significant DM decrease by roughly 5 pc cm$^{-3}$ yr$^{-1}$ \citep{snelders25}, arguing strongly for a complex, highly magnetized immediate FRB environment. Additionally, \cite{ayush_2026} show FRB 20220529A exhibits a DM variation of $-0.881 \pm 0.001$ pc cm$^{-3}$ yr$^{-1}$ accompanied by RM variations of tens of rad m$^2$. The contrasting DM and RM evolution observed in \tns compared to other repeaters sets it apart and implies a broad range of conditions in the immediate environment of repeating FRBs. This diversity may indicate the presence of multiple source classes, even within the population of hyperactive repeaters.

\subsection{Total Energy Budget}\label{Discussion/Total Energy Budget}
From their observation of over 1000 bursts from \tns using FAST, \cite{zhang_2023_fast_observations_r117} suggest that the total energy emitted from the source in the radio band might be larger than is plausible from some models invoking a magnetar as the central engine.
Here we estimate, on the basis of our CHIME observations, the total energy emitted by the source over our 1.5 years of observations and compare it to the total dipolar energy in a typical magnetar's external magnetic field, to similarly consider whether a magnetar model is plausible for the source.

First, we calculate the energy emitted solely through the FRB emission that we have observed, using Equation (1) from \cite{zhang_2023_fast_observations_r117}:
\begin{equation}\label{eq: energy}
    E_{FRBs} = 10^{39} \text{erg}\sum_{n=1}^N \left(\frac{4\pi \cdot D_L^2}{(1+z)^2(10^{28}\text{cm})^2}\right) \left(\frac{F_n}{\text{Jy}\cdot\text{ms}}\cdot\frac{\text{BW}_n}{\text{GHz}}\right)
\end{equation}
\noindent where $D_L$ is the luminosity distance to the source, 362.4(1) Mpc \citep{r117_host_galaxy}, $F_n$ is the fluence of the $n$\textsuperscript{th} burst, $N$ is the total number of bursts we have observed (\numfbfits in this study, as we require robust \texttt{fitburst} models to measure the bandwidth\footnote{For some bursts, \texttt{fitburst} converged to an incorrect solution, so we say ``robust" here to indicate that \texttt{fitburst} converged to a solution with no obvious structure in the residuals.}), BW$_n$ is the $n$\textsuperscript{th} burst's bandwidth, and $z$ is the redshift to the source, 0.0771. In this expression, the first factor in parentheses in the summation follows from the fact that this equation assumes isotropic emission and the second factor in parentheses in the summation represents the energy observed by the telescope. 

The assumption of isotropic emission could result in an overestimation of the total energy emitted, as beamed radiation could result in the same observed fluence with much smaller intrinsic energy. To account for this, we multiply Equation~\ref{eq: energy} by a beaming factor, $f_b = \delta\Omega / 4\pi$. 
Although this quantity is unknown, we use a value of 0.1 as in \cite{fast_20201124A_extreme_activity} and \cite{beaming_factor}.

Since radiation in the radio band does not represent the entirety of the emitted energy, we scale the observed radio energy by a radio efficiency factor, $\eta_r = F_r / F_{tot}$ where $F_r$ is the radio fluence and $F_{tot}$ is the fluence across all other wavelengths. The value of $\eta_r$ is poorly constrained for the FRB population, as only the Galactic magnetar SGR 1935+2154 (FRB 20200428D), has been observed to emit both FRB-like radio bursts and high energy emission (\citealt{bochenek_sgr}, \citealt{chimefrb_sgr}). From observations of SGR 1935+2154, the efficiency factor between radio and X-ray fluence was estimated to be between 10$^{-4}$ and 10$^{-5}$ (\citealt{eta_lower}, \citealt{eta_upper}). It is important to note that the radio to X-ray fluence ratio must be less than the radio efficiency factor, as it excludes energy emitted at wavelengths other than X-rays. For \tns, \cite{amanda_xray_followup} placed a lower limit on the radio to X-ray fluence ratio of $10^{-7}$ through contemporaneous X-ray observations. In this analysis, we conservatively use an efficiency factor value of 10$^{-4}$, but note that this value could be much lower, which for a given radio energy, would imply a higher total energy. 

Finally, we account for the observing duty cycle, $\zeta = t_{obs} / t_{tot}$ where $t_{obs}$ is the total observing time and $t_{tot}$ is the total elapsed time throughout the observations. This term must be accounted for as the CHIME/Pulsar instrument is not observing the source at all times, thus misses a large portion of the emitted bursts. The total exposure in this study was 201.2 hours over 554 days, implying $\zeta = 0.015$. 

The total energy emitted by \tns given the CHIME observations is thus estimated as:

\begin{equation}\label{eq: total_energy_emitted}
    E_{tot} = 10^{43}\text{erg}
    \left(\frac{E_{FRBs}}{10^{39}\text{erg}}\right)
    \left(\frac{f_b}{0.1}\right)
    \left(\frac{\eta_r}{10^{-4}}\right)^{-1}
    \left(\frac{\zeta}{0.015}\right)^{-1}.
\end{equation}

Using our observed burst properties, the estimated total energy required is $E_{tot} = 2 \times 10^{43}$ ergs. With a less conservative efficiency factor ($\eta_r = 10^{-7}$), $E_{tot}$ could be as high as $2 \times 10^{46}$ ergs. Furthermore, this does not account for the 284 FRB candidates that do not have robust \texttt{fitburst} models. This is comparable and somewhat higher than estimated by \cite{fast_20201124A_extreme_activity} for their observations during the initial high-rate activity period.

Based on the properties of Galactic magnetars \citep{magnetar_cat}, we take the typical dipolar surface magnetic field strength of a magnetar to be 10$^{14}$ G. Assuming a radius of $10$ km,  we estimate the magnetic energy stored in  a magnetar to be {$E_{mag.} = 2 \times 10^{45}\text{erg}[1/6(B/10^{14}\text{G})^2(R^3 / 10^{6}\text{cm})^3]$. Comparing this with the estimated energy emitted by the FRB source, the latter over our 1.5 years of observation amounts to  $\sim$1\% of the total magnetic energy of a 10$^{14}$ G magnetar, as argued by \cite{fast_20201124A_extreme_activity}, and is {\it higher} than this estimated total magnetar energy for less conservative assumptions.

However, some magnetars have estimated dipolar field strength well in excess of $10^{14}$~G (e.g., SGR 1900+14, 1E~1841$-$045), even as high as $2 \times 10^{15}$~G in SGR 1806$-$20 \citep[see][for a compendium]{magnetar_cat}. Such fields represent a lower limit on the true field since they are estimated from spin down, which is sensitive to the surface dipole component only, whereas magnetars may well have higher-order multipoles and/or much higher internal fields.
For example, SGR~0418+5729 has a surface dipolar field strength of only $7.5 \times 10^{12}$ G despite of showing clear magnetar phenomenology \citep{ret+10}, suggesting a much higher internal field. In principle, a neutron star may be able to carry fields as high as $10^{17}-10^{18}$~G \citep[e.g.,][]{ls91,mgt+11}.
Given the extreme natures of highly active repeaters like \tns, extreme forms of magnetars, particularly at birth, seem plausible and would alleviate the energy budget problem.

Nevertheless, these models continued to be challenged with new discoveries such as that by \cite{omar-2026-max-energy} which shows that rare, high-energy bursts can contain as much energy as the hundreds or even thousands of lower-energy bursts. So the source of FRBs also needs to be able to liberate large amounts of energy in a very short amount of time. In addition, new discoveries of sources with even higher activity rates and emitted energy than \tns (e.g., \citealt{fast_20201124A_extreme_activity}, \citealt{meerkat_20240619d}, and \citealt{zhang2025prolificrepeatingfastradio}), energy analyses like the above will continue to provide lower limits on the total energy budget of sources of hyperactive repeating FRBs, or even rule out certain magnetar models, such as those with large beaming factors and low efficiency factors  \citep[e.g.][]{synchrotron_maser}.

% Conclusion
\section{Conclusion}\label{sec: Conclusion}
We have presented the search for and analysis of \numbursts FRBs from the hyperactive repeater \tns. We describe an FRB search pipeline and a burst property pipeline designed for the CHIME/Pulsar instrument and explain the advantages of this pipeline over CHIME/FRB for dedicated follow-up of repeating FRBs. We present analyses of the burst parameters as measured by \texttt{fitburst} for \numfbfits of the detected sample of FRBs. We show that the wait-time distribution is bimodal and that the DM increases linearly with time in the CHIME band over our observing duration, at a rate of \dmvar, while the source's RM remains near zero, arguing against a highly magnetized local medium for the source, in contrast to what has been observed for other active repeaters.  We estimate the total energy output of the source from our detections to be $E_{tot} > 2 \times 10^{43}$ ergs and argue that extreme magnetar models invoking strong dipoles, high-order surface multipoles, or high internal fields can provide sufficient energy.  
We encourage long-term follow-up observations of repeating FRBs in hyperactive states to further constrain FRB progenitor models.

\section{Acknowledgments}
\begin{acknowledgments}
T.C.A is supported by the Centre de recherche en astrophysique du Québec, un regroupement stratégique du FRQNT. A.B.P. acknowledges support by NASA through the NASA Hubble Fellowship grant HST-HF2-51584.001-A awarded by the Space Telescope Science Institute, which is operated by the Association of Universities for Research in Astronomy, Inc., under NASA contract NAS5-26555. A.B.P. also acknowledges prior support from a Banting Fellowship, a McGill Space Institute~(MSI) Fellowship, and a Fonds de Recherche du Quebec -- Nature et Technologies~(FRQNT) Postdoctoral Fellowship. V.M.K. holds the Lorne Trottier Chair in Astrophysics \& Cosmology, a Distinguished James McGill Professorship, a Lowy Visiting Professorship at Tel Aviv University, and receives support from an NSERC Discovery grant (RGPIN 228738-13). A.P. is a Trottier Space Institute Postdoctoral Fellow. A.M.C. is a Banting Postdoctoral Researcher. A.P.C. is a Canadian SKA Scientist and is funded by the Government of Canada / est financé par le gouvernement du Canada. E.F. is supported by the National Science Foundation under grant AST-2407399. DCG is supported by NSF AST Award 2406919. The AstroFlash research group at McGill University, University of Amsterdam, ASTRON, and JIVE is supported by: a Canada Excellence Research Chair in Transient Astrophysics (CERC-2022-00009); an Advanced Grant from the European Research Council (ERC) under the European Union's Horizon 2020 research and innovation programme (`EuroFlash'; Grant agreement No. 101098079); an NWO-Vici grant (`AstroFlash'; VI.C.192.045); an ERC Starting Grant (`EnviroFlash'; Grant agreement No. 101223057); and an NWO-Veni grant (VI.Veni.222.295). C. L. acknowledges support from the Miller Institute for Basic Research at UC Berkeley. K.N. acknowledges support by NASA through the NASA Hubble Fellowship grant \# HST-HF2-51582.001-A awarded by the Space Telescope Science Institute, which is operated by the Association of Universities for Research in Astronomy, Incorporated, under NASA contract NAS5-26555. M.N. is a Fonds de Recherche du Quebec - Nature et Technologies (FRQNT) postdoctoral fellow. The AstroFlash research group at McGill University, University of Amsterdam, ASTRON, and JIVE is supported by: a Canada Excellence Research Chair in Transient Astrophysics (CERC-2022-00009); an Advanced Grant from the European Research Council (ERC) under the European Union's Horizon 2020 research and innovation programme (`EuroFlash'; Grant agreement No. 101098079); an NWO-Vici grant (`AstroFlash'; VI.C.192.045); an ERC Starting Grant (`EnviroFlash'; Grant agreement No. 101223057); and an NWO-Veni grant (VI.Veni.222.295). P.S. acknowledges the support of an NSERC Discovery Grant (RGPIN-2024-06266). V.S. is supported by a Fonds de Recherche du Quebec - Nature et Technologies (FRQNT) Doctoral Research Award.
\end{acknowledgments}

\appendix
\section{Fitburst Fit Examples}\label{sec: appendix/fitburst_fits}
\begin{figure}[h]
    \centering
    \includegraphics[width=\linewidth]{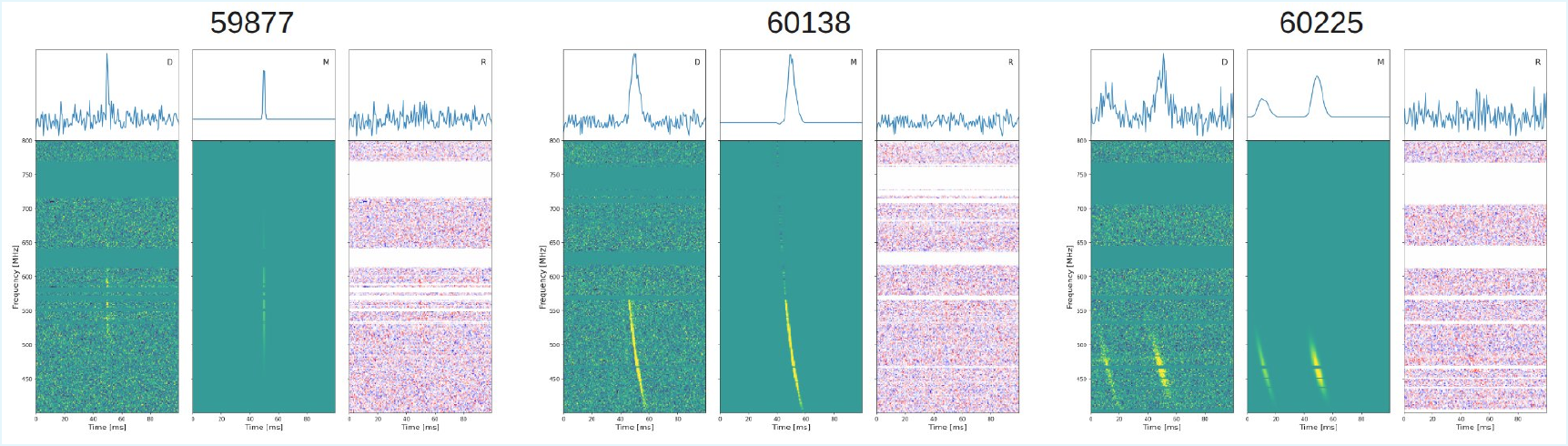}
    \caption{Example of three \texttt{fitburst} fits with bursts detected on MJD 59877, 60138, and 60225, from left to right. Each panel shows the burst's dynamic spectrum (left panel), the \texttt{fitburst} model (middle panel), and the residual (right panel). The dynamic spectra are all dedispersed to the detection DM ($219.456$ pc cm$^{-3}$) causing the dispersive sweep to be evident at later MJDs.}
\end{figure}
% \section{Calculation of Upper Limit on the Magnetic Field in the Source Environment}
\label{appendix:A}

% \subsection{Upper limit on the magnetic field strength in the FRB's environment: }\label{sec: appendix/upper-limit-on-B}
% In general, the RM is given by: 
% \begin{equation}
%     \text{RM} = \frac{e^3}{2\pi m_e^2 c^4}\int_{LOS} n_e B_{\parallel}dl
% \end{equation}

% In particular, within the region of the line of sight where the DM variation is occurring, we have:

% \begin{equation}
%     \text{RM}_{env} = \frac{e^3}{2\pi m_e^2 c^4}\text{B}_{\parallel, env}\int_{env} n_e dl
% \end{equation}

% \begin{equation}
%     \text{RM}_{env} = \frac{e^3}{2\pi m_e^2 c^4}\text{B}_{\parallel, env} \text{ DM}_{env}
% \end{equation}

% Taking the time derivative yields: 

% \begin{equation}\label{eq: rm}
%     \frac{d\text{RM}_{env}}{dt} = \frac{e^3}{2\pi m_e^2 c^4}\text{B}_{\parallel, env} \frac{d\text{ DM}_{env}}{dt}
% \end{equation}

% \begin{equation}
%     \text{B}_{\parallel, env}  =  \frac{e^3}{2\pi m_e^2 c^4} \frac{d\text{ DM}_{env}}{dt}  \left(\frac{d\text{RM}_{env}}{dt}\right)^{-1}
% \end{equation}

% Using values of 1.4 pc cm$^{-3}$ yr$^{-1}$ and the $3\sigma$ upper limit on the RM variation of 13.4  rad m$^{-2}$, we obtain a $3\sigma$ upper limit on the parallel component of the magnetic field of 8.8 $\mu$G. 

\bibliography{references}{}
\bibliographystyle{aasjournal}

\end{document}